\documentclass[aps,prd,twocolumn,floatfix]{revtex4}
\usepackage{amsfonts}
\usepackage{amssymb}
\usepackage{amsmath}
\usepackage{graphicx}
\usepackage{graphics}
\usepackage{color}
\usepackage{epsfig}
\newcommand{\be}{\begin{equation}}
\newcommand{\ee}{\end{equation}}
\newcommand{\bea}{\begin{eqnarray}}
\newcommand{\eea}{\end{eqnarray}}
\newcommand{\bwt}{\begin{widetext}}
\newcommand{\ewt}{\end{widetext}}
\begin{document}
\title{On Local Dilatation Invariance}
\author{T.E. Clark}
\email[e-mail address:]{clarkt@purdue.edu}
\affiliation{Department of Physics,\\
 Purdue University,\\
 West Lafayette, IN 47907-2036, U.S.A.}
\author{S.T. Love}
\email[e-mail address:]{loves@purdue.edu}
\affiliation{Department of Physics,\\
 Purdue University,\\
 West Lafayette, IN 47907-2036, U.S.A.}
\begin{abstract}
{The relationship between local Weyl scaling invariant models and local dilatation invariant actions is critically scrutinized. While actions invariant under local Weyl scalings can be constructed in a straightforward manner, actions invariant under local dilatation transformations can only be achieved  in a very restrictive case. The invariant couplings of matter fields to an Abelian vector field carrying a non-trivial scaling weight can be easily built, but an invariant Abelian vector kinetic term can only be realized when the local scale symmetry is spontaneously broken.}
\end{abstract}

\maketitle

Actions invariant under rigid scaling or dilatation transformations can be converted into actions which exhibit a rigid Weyl invariance \cite{OR}. When the Weyl transformations are promoted to become local symmetries, invariant actions can be constructed by gauging in the familiar fashion \cite{OR}. On the other hand, to our knowledge, there has been no analysis of the gauging of the dilatation symmetry for four dimensional Lorentz invariant quantum field theories.  \\

Consider the uniform, rigid space-time scaling or dilatation transformation 
\bea
x^\mu\rightarrow x^{\prime\mu}=e^{\omega}x^{\mu} ,
\eea
where $\omega$ is a space-time independent parameter. Under such a transformation, a local field, $\Delta(x)$ is said to transform with scaling weight $d_\Delta$ provided
\bea
\Delta(x)\rightarrow e^{d_\Delta\omega}\Delta(x^\prime) .
\eea
Note that the space-time point scaling is accompanied by a field rescaling so this is not just a special case of a general coordinate transformation. Classically invariant models in four space-time dimensions are constructed under such transformations provided each action term has scaling weight four. For example, if $\phi(x)$ is a scalar field with scaling weight one, a rigid dilatation invariant action is 
\bea
S=\int d^4x [\frac{1}{2}\partial_\mu \phi(x)\partial^\mu\phi(x)-\frac{\lambda}{4}\phi^4(x)].
\eea

An alternate approach to discuss scaling is to keep the space-time point fixed but introduce a background space-time metric tensor $g^{\mu\nu}(x)$ which compensates for the space-time scaling. Under this rigid Weyl transformation, a field  $\Delta(x)$ has scaling weight $d_\Delta$ provided it transforms as
\bea
\Delta(x)\rightarrow e^{d_\Delta\omega}\Delta(x).
\eea
Note that contrary to the rigid dilatation transformation, this transformation leaves space-time points unchanged. The non-dynamical metric tensor is defined to transform with scaling weight two so that
\bea
g^{\mu\nu}(x)\rightarrow e^{2\omega}g^{\mu\nu}(x).
\eea
Invariant actions can then be constructed as
\bea
S&=&\int d^4x \sqrt{-g}{\cal O}(x),
\eea
where each operator ${\cal O}(x)$ has Weyl scaling weight four. Once again, if we consider a scalar field, $\phi(x)$, with Weyl scaling weight one, a rigid Weyl invariant action is
\bea
S=\int d^4x \sqrt{-g} [\frac{1}{2}\partial_\mu \phi(x)g^{\mu\nu}(x)\partial_\nu\phi(x)-\frac{\lambda}{4}\phi^4(x)].
\eea

Now consider allowing the transformation parameter $\omega$ to be a function of space-time: $\omega(x)$. Under such local Weyl transformations, the field $\Delta(x)$ has scaling weight $d_\Delta$ and transforms as
\bea
\Delta(x)\rightarrow e^{d_\Delta\omega(x)}\Delta(x).
\eea
Note that the space-time point is still unchanged.  
To make invariants, proceed just as in the rigid case, with the replacement of $\partial_\mu\Delta(x)$ by $D_\mu\Delta(x)=(\partial_\mu -d_\Delta A_\mu(x))\Delta(x)$ where under the local Weyl scaling, the vector field $A_\mu(x)$ has Weyl scaling weight zero:
\bea
A_\mu(x)\rightarrow A_\mu(x)+\partial_\mu\omega(x).
\eea
In that case, under the local Weyl scalings, the covariant derivative of $\Delta(x)$ also has Weyl scaling weight $d_\Delta$:
\bea
D_\mu\Delta(x)\rightarrow e^{d_\Delta\omega(x)}D_\mu\Delta(x).
\eea
Since the metric tensor continues to transform with Weyl scaling weight two
\bea
g^{\mu\nu}(x)\rightarrow e^{2\omega(x)}g^{\mu\nu}(x),
\eea
the local Weyl scaling invariant action for a scalar field, $\phi(x)$, with local Weyl scaling weight one is readily constructed as
\bea
S=\int d^4x \sqrt{-g}[\frac{1}{2}D_\mu\phi(x) g^{\mu\nu}(x)D_\nu\phi(x)-\frac{\lambda}{4}\phi^4(x)].
\eea
The locally Weyl invariant kinetic term for the vector is made using the field strength
\bea
F_{\mu\nu}(x)=\partial_\mu A_\nu(x)-\partial_\nu A_\mu(x)
\eea
which transforms with Weyl scaling weight zero
\bea
F_{\mu\nu}(x)\rightarrow F_{\mu\nu}(x).
\eea
Said action is then simply given by
\bea
S &=&\int d^4x \sqrt{-g} [-\frac{1}{4}F_{\mu\nu}(x)g^{\mu\lambda}(x)g^{\nu\rho}(x)F_{\lambda\rho}(x)].
\eea

Now consider a local dilatation transformation where the transformed space-time point, $x^{\prime\mu}$, is implicitly defined as
\bea
\frac{\partial x^{\prime \mu}}{\partial x^\nu}=e^{\omega(x)}\eta^\mu_\nu .
\eea
Note that this reduces to the rigid scale transformation: $\omega(x)=\omega$ and $x^{\prime \mu}=e^{\omega}x^\mu$ when the parameter $\omega$ is independent of $x^\mu$. 
A local field, $\Delta(x)$, has local scaling weight $d_\Delta$ provided it transforms as
\bea
\Delta(x)\rightarrow e^{d_\Delta\omega(x)}\Delta(x^\prime).
\eea
Once again, there is a field local rescaling accompanying the space-time point local rescaling. Thus this is not simply a general coordinate transformation and invariants cannot be constructed using the methods of general relativity. 
Defining the covariant derivative of $\Delta(x)$ as 
\bea
D_\mu\Delta(x)\equiv (\partial_\mu -d_\Delta V_\mu(x))\Delta(x)
\eea
where $V_\mu(x)$ has the local scaling transformation law
\bea
V_\mu(x)\rightarrow e^{\omega(x)}V_\mu(x^\prime)+\partial_\mu \omega(x),
\label{Vmu}
\eea
it then follows that
\bea
D_\mu\Delta(x)&\rightarrow & e^{(d_\Delta+1)\omega(x)}D^\prime_\mu\Delta(x^\prime)
\eea
transforms with scaling weight $d_\Delta +1$. Clearly, 
$D_\mu D_\nu\Delta(x)=\left(\partial_\mu-(d_\Delta +1)V_\mu(x)\right)\left(D_\nu\Delta(x)\right)$ transforms with scaling weight $d_\Delta +2$. Note that  $V_\mu(x)$ transforms with dilatation scaling weight one in addition to the inhomogeneous piece. Recall that the Weyl scaling weight of the analogous vector is zero. 
Local dilatation invariants can then be constructed out of terms with dilatation scaling weight four. Thus for a scalar field with local scaling weight one, we construct the locally invariant action 
\bea
S&=&\int d^4x [\frac{1}{2}\left(D_\mu\phi(x)\right)\left(D^\mu\phi(x)\right)-\frac{\lambda}{4}\phi^4(x)].
\eea

Now consider a kinetic term for the vector field $V_\mu(x)$.  
A first approach is to introduce the field strength $F_{\mu\nu}(x)=\partial_\mu V_\nu (x)-\partial_\nu V_\mu (x)$. However, under local dilatations, this field strength transforms as
\bea
&&F_{\mu\nu}(x)
\rightarrow e^{2\omega (x)} F_{\mu\nu}(x^\prime) \cr
&&~~+e^{\omega(x)}(\partial_\mu \omega(x) V_\nu(x^\prime)-\partial_\nu \omega(x) V_\mu(x^\prime))
\eea
which is not an object with scaling weight two. Thus the action $\int d^4x F_{\mu\nu}(x)F^{\mu\nu}(x)$ is not an invariant and the construction of the massless vector kinetic term is not at all straightforward. In fact it is impossible. This is a consequence of the Weinberg-Witten theorem \cite{WW} which forbids the appearance of a massless vector carrying a non-trivial charge (in this case, a scaling weight) of a Lorentz covariant conserved current (in this case, the dilatation current). Recall that the massless vector appearing in the gauging of the Weyl symmetry does not carry a Weyl scaling weight.

One possible realization of the scaling is as a spontaneously broken symmetry. To do so, we introduce the novel scalar degree of freedom $\sigma(x)$ and define the operator $e^{\sigma(x)}$ to have scaling weight one so that
\bea
e^{\sigma(x)}\rightarrow e^{\omega(x)}e^{\sigma(x^\prime)}.
\eea
It follows that
\bea
\sigma(x)\rightarrow \sigma(x^\prime)+\omega(x).
\eea
Thus $\sigma(x)$ is a dilaton \cite{CLD1}-\cite{CLD2} and the scale symmetry is being realized a la Nambu-Goldstone. A locally scale invariant kinetic term for $V_\mu(x)$ is now readily constructed as
\bea
S&=&-\frac{1}{4}\int d^4x (e^{-\sigma(x)}[D_\mu, D_\nu]e^{\sigma(x)})^2 \cr
&=&-\frac{1}{4}\int d^4x [(\partial_\mu V_\nu (x) -\partial_\nu V_\mu (x))^2 +...]
\eea
where 
\bea
D_\mu e^{\sigma(x)} =(\partial_\mu -V_\mu(x)) e^{\sigma(x)}= e^{\sigma(x)}\left(\partial_\mu\sigma(x)-V_\mu(x)\right).
\eea
The locally scale invariant kinetic term for the dilaton is
\bea
S&=& \frac{1}{2}\int d^4x \left(D_\mu e^{\sigma(x)}\right)\left(D^\mu e^{\sigma(x)}\right)
\eea
which contains in it a mass term for the vector: i.e. the Higgs mechanism is operational.\\
\\
Thus it seems a model with spontaneously broken local scale invariance can be constructed. 
 One clue pertaining to this highly unusual behavior may lie in the Abelian vector field transformation law:
\bea
V_\mu(x)\rightarrow V_\mu^\prime(x)=e^{\omega(x)}V_\mu(x^\prime)+\partial_\mu\omega(x).
\eea
Recall that a massless Abelian vector is purely transverse. With above transformation law, both the transverse and longitudinal components have gauge freedom. It follows that it cannot describe the transformation law of a massless vector. Hence the only way to realize the symmetry is for the vector to develop a mass term a la the Higgs mechanism. 
Further note that, in general, the scale symmetry will be anomalous \cite{CJ} and thus it cannot be gauged without violating unitarity. The presence of a dilaton allows for a Wess-Zumino term \cite{WZ} which can restore local symmetry.

\begin{acknowledgments}
We thank Tonnis ter Veldhuis for enjoyable discussions. The work of TEC and STL was supported in part by the U.S. Department of Energy under grant DE-FG02-91ER40681 (Theory).  
\end{acknowledgments}

\end{document}